\documentclass[aps,prx,superscriptaddress,twocolumn,floatfix,citeautoscript]{revtex4-1}

\usepackage{graphicx}
\usepackage{color}
\usepackage{float}
\usepackage{amsmath}
\usepackage{subfigure}
\usepackage{multirow}
\usepackage{appendix}

\newcommand{\CA}[1]{\textcolor{blue}{#1 }}

\newcommand{\CHALKRIVER}{Canadian Nuclear Laboratories, Chalk River, Ontario, Canada}
\newcommand{\QUEENS}{Department of Mechanical and Materials Engineering, Queen's University, Kingston, Ontario, Canada}
\newcommand{\INL}{Idaho National Laboratory, Idaho Falls, Idaho, ID 83415}

\begin{document}

\title{Automated \CA{Detection}of Helium Bubbles in Irradiated X-750}

\author{Chris M. Anderson}
\affiliation{\QUEENS}

\author{Jacob Klein}
\affiliation{\QUEENS}

\author{Heygaan Rajakumar}
\affiliation{\CHALKRIVER}

\author{Colin D. Judge}
\affiliation{\INL}

\author{Laurent Karim B\'{e}land}
\affiliation{\QUEENS}
\email[]{laurent.beland@queensu.ca}

\date{\today}

\begin{abstract}
Imaging nanoscale features using transmission electron microscopy is key to predicting and assessing the mechanical behavior of structural materials in nuclear reactors. Analyzing these micrographs is often a tedious and labour intensive manual process. It is a prime candidate for automation. Here, a region-based convolutional neural network is adapted to detect helium bubbles in micrographs of neutron-irradiated Inconel X-750 reactor spacer springs. We demonstrate that this neural network produces analyses of similar accuracy and reproducibility to that produced by humans. Further, we show this method as being four orders of magnitude faster than manual analysis allowing for generation of significant quantities of data. The proposed method can be used with micrographs of different Fresnel contrasts and magnification levels.

\end{abstract}

\maketitle

\section{Introduction}
Transmission electron microscopy (TEM) enables microstructural characterization of materials with nanoscale precision; this methodology is now ubiquitous in materials science \cite{jenkins2000characterisation,zhang2014tem}. TEM imaging allows insights into microstructural behaviour and defect morphology at nanometer scales, ultimately leading to knowledge which can be translated at the system level. TEM is utilized frequently by the nuclear power industry as a  visualisation tool for irradiation damage. It is also used by scientists to identify material degradation in advance of component failure \cite{jenkins2000characterisation}. The ability to image ex-situ components and visualize their microstructures is important to predict a material's response to irradiation \cite{zhang2014tem}.

Irradiation-induced damage of reactor components is one of the primary issues that plagues nuclear power generation due to the high cost of component replacement and the introduction of uncertainty into life cycle predictions. Neutrons interact with the atoms through numerous mechanisms resulting in a multitude of defect types, each with unique consequences towards macroscopic behaviour of the system. An issue of particular interest is the build-up of nanoscale helium (He) bubbles in nickel-based superalloys used in Canada Deuterium Uranium (CANDU) reactors, namely Inconel X-750 \cite{judge2015intergranular,griffiths2014effect}. The helium is mainly produced through the interactions of thermal neutrons emitted by the reactor core with nickel (Ni) atoms when a collision occurs. This reaction is the transmutation of $^{59}$Ni to $^4$He and $^{56}$Fe through the absorption of a neutron \cite{judge2015effects}. This interaction probability is governed by the thermal neutron cross section \cite{Block2010}. $^{59}$Ni has a high thermal cross section of 1000 Barns, this cross section leads to increased helium production from the high interaction probability \cite{Kopecky1997}. The presence of helium has important effects on the mechanical properties of the structural alloy. Helium first coalesces into bubbles, which act as nucleation points for voids, this in turn can lead to swelling. Helium accumulation is also thought to be linked to grain boundary embrittlement \cite{judge2015effects}. Under typical reactor-operation conditions, the sizes of helium bubbles in Ni-based superalloys are generally less than 10 nanometers \cite{snoeck2006characterization}. The small bubble sizes generated require the use of high contrast imaging to monitor the effects of bubble ingress.

Helium bubbles appear as circular objects in TEM micrographs under Fresnel contrast imaging \cite{judge2015intergranular}. When imaging bubbles, bright field techniques are often used to generate the required contrast to visualize defects less than 5 nm in diameter \cite{zheng2019microstructure}. The use of bright field is not ideal when quantifying defects as the bubble fringes produced through bright field are only an approximation of the true edge. The images produced by TEM suffer from noise due to irradiation and 2D projections of bubbles can overlap. This leads to a time consuming and error prone manual quantification process given the lack of appropriate software tools to automate the classification process \cite{judge2018high}. Manual quantification of the images creates a large bottleneck in the quantification of the structural degradation occurring in components. There are three main downsides to manual identification of bubbles. First, the process is time consuming: an individual image can take up to a few hours to classify. Second, manual identification is error-prone as bubbles can be easily misidentified. Third, there is a lack of reproducibility and consistency from one human inspection to another. 

Advances in image recognition algorithms have led to recent adoptions in numerous fields. Cirecsan \textit{et al.} apply these algorithms in their pioneer report, in which they describe using a convolutional neural network (CNN) to detect cell mitosis associated to breast cancer \cite{cirecsan2013mitosis}. This work highlighted the viability of neural networks as image classifiers with moderate computational costs. CNNs are now able to identify defects in  TEM images, including noisy TEM images \cite{zhu2017deep}. Recently, a method combining fast-Fourier transforms with CNNs proved an efficient method for identification of phase transformations in Tungsten disulfide (WS$_2$) characterized by TEM and scanning tunnelling microscopy \cite{vasudevan2018mapping,maksov2019deep}. Object detection was also used to detect dislocation loops in irradiated FeCrAl alloys, with successes in the extraction of both visual and quantitative defect metrics matching manual methods \cite{li2018automated}. We also note the recent work of Roberts \textit{et al.}, in which they developed a method to identify common crystallographic details using a hybrid network architecture and semantic segmentation. The method can extract information from micrographs with a high density of features, netting a stark improvement over time-demanding and error-prone manual quantification \cite{roberts2019deep}. Recent developments in the field of object detection, namely region proposal methods (R-CNNs), suggest that they are viable analysis methods for TEM images.

In this paper, we adapt Faster R-CNN \cite{NIPS2015_5638} to automatically identify helium bubbles in irradiated X-750 micrographs by adjusting its hyper-parameters and introducing an image preparation procedure. First, the network architecture is introduced. Second the preparation of the training and validation data is discussed. Third the validation metrics of the network are described and the performance of the model is assessed. Finally, prescriptions for use of this algorithm are made.

\section{Methodology}

\subsection{Strategy overview}

We opted for a two-stage detector based on the Faster R-CNN architecture developed by Ren \textit{et al.} \cite{NIPS2015_5638}. The model is trained and validated using micrographs generated and manually analysed by the Canadian Nuclear Laboratories (CNL) in the context of its commercial activities involving X-750 spacer samples extracted from the CANDU nuclear reactor fleet. The model aims at identifying the location of bubbles, their radii, and their cumulative volumes.

\subsection{Network Architecture}
\label{sec:NetArch}
 Historically, CNNs have been hindered by their high computational cost. Mitigation strategies include the sharing of convolutional layers across both region proposals and region detection networks, which leads to significant reductions in computational cost \cite{Girshick_2015_ICCV,NIPS2015_5638}. The R-CNN differs from the CNN, in that the two-stage detection model generates region proposals which are then passed forward to the convolutional layers and used to narrow the search for objects of interest \cite{girshick2014rich}. The use of region proposals alone are not enough to make R-CNNs effective for rapid classification but through the sharing of layers they have become viable means for fast and effective object detection models \cite{Girshick_2015_ICCV,NIPS2015_5638}. Recent efforts led to the development of Faster R-CNN, which is pre-trained on the COCO dataset and publicly available. This model offers a balance between base mean average precision (mAP) on the COCO dataset and computational costs. Faster R-CNN models utilize a deep internal network to generate a feature network off of a base feature map much like the Fast R-CNN methodology \cite{NIPS2015_5638}. The feature network is then passed to a region proposal network (RPN), which shares its layers with the last convolutional layers. The RPN is then used to generate the regions of interest (RoI) where objects are expeced to be located. The model then utilizes the feature map along with the RoIs to make object location predictions utilizing the same detection model as Fast R-CNN \cite{Girshick_2015_ICCV}. Deriving the region proposals from the feature maps using shared convolutional layers reduces the time required to classify an image without sacrificing accuracy \cite{NIPS2015_5638}. The RPN, utilizing the feature map to generate its region proposals, makes the Faster R-CNN an ideal candidate for the detection of a large quantity of small bubble defects.
 
\begin{figure}
\centering
\includegraphics[width=8.6cm]{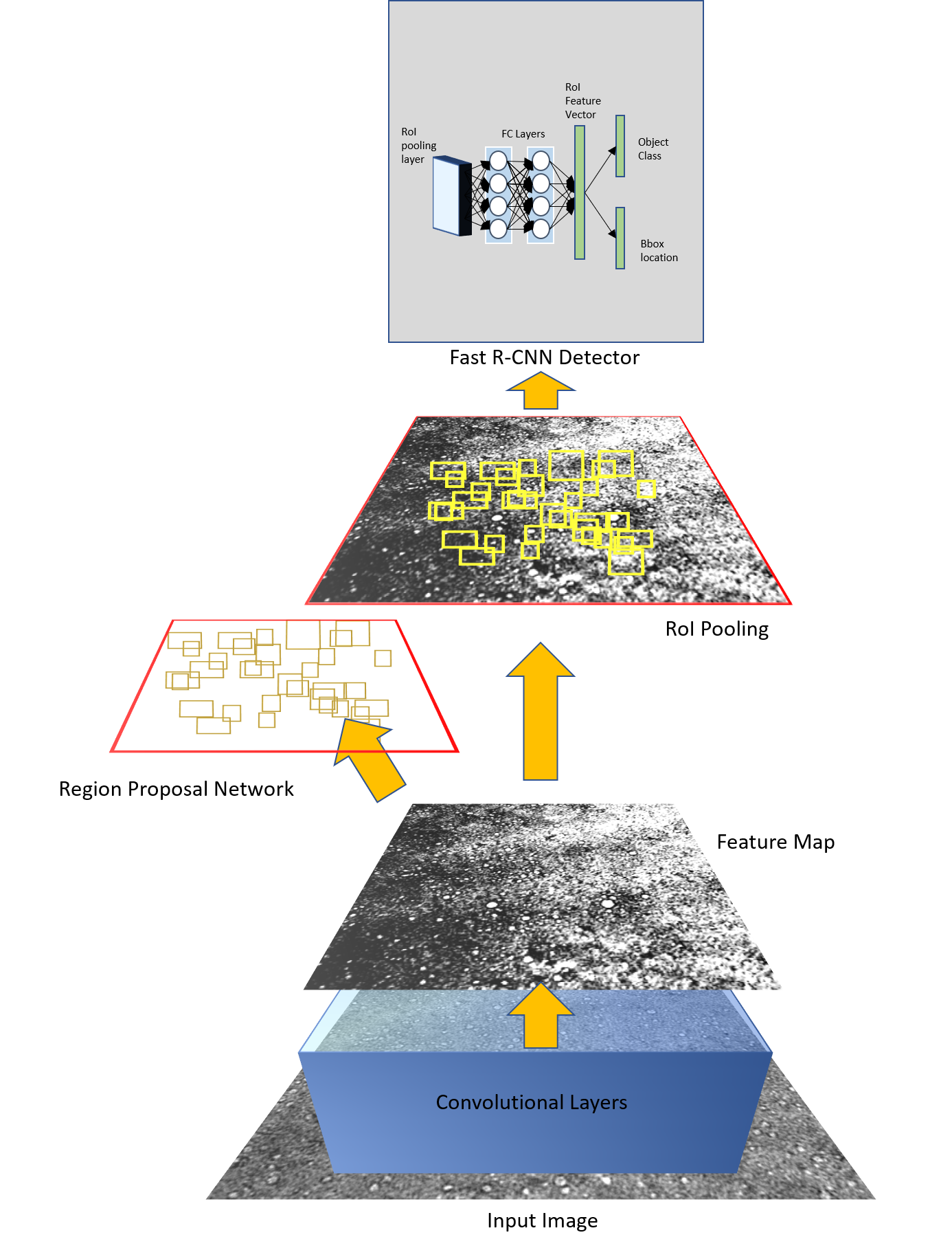}
\caption{An illustration of the Faster R-CNN architecture. A collective feature map is generated for the entire image which the region proposal network then draws from to generate the RoIs. Each region is then overlaid on the original feature map and are then passed to the Fast R-CNN detector network for individual region classifications.}
\label{fig:rcnn}
\end{figure}

 The architecture of the network used in this paper is the single unified Faster R-CNN which consists of a region proposal network and a detection network. The input layer accepts a standard image with no dimensional requirements, the input is then passed forward to the convolutional layers where a feature map is generated. The as processed feature map is then passed to the RPN sharing the final convolutional layers. The layers utilize a sliding window detector which moves across the feature map taking an input as a $n \times n$ window. Each window is then fed to parallel fully-connected layers, a box classification and box regression layer which generate the bounding box location and objectness score \cite{NIPS2015_5638}. At each sliding window location multiple region proposals are predicted with a limit on maximum predictions denoted as the tuneable value $k$. The feature map and outputs are then passed to the final detection module which is carried forward from the Fast R-CNN methodology and classifies each region individually. An overview of the unified network structure is shown in Fig. \ref{fig:rcnn} where the input image is fed forward and the resultant feature map is used to detect object classes. The output of the Faster R-CNN is an annotated .jpeg image with detected objects enclosed within bounding boxes. The coordinate values of these bounding boxes are preserved in a .csv file.
 
\subsection{Data Collection}

The data was collected as part of an effort to characterize the helium ingress in X-750 reactor spacer springs in the CANDU reactor fleet. The nominal composition of the current Inconel X-750 alloy is summarized in Table \ref{tab:comp}. Ideally, one would perform high-angle annular dark-field scanning transmission electron microscopy (HAADF-STEM) to measure bubble size with high accuracy. However, this imaging technique is typically not possible when imaging high densities of cavities less than 5 nm in diameter \cite{zheng2019microstructure}. As such, bright field TEM imaging was performed, samples were prepared from an ex-service spacer spring.  The maximum flux of fast neutrons emitted from the CANDU fuel bundles is $3.5\cdot10^{17} nm^{-2}s^{-1}, E > 1$ MeV \cite{jenkins2000characterisation}. For a typical CANDU fuel channel power profile, each Ni atom will be displaced approximately once per year by fast neutrons \cite{jenkins2000characterisation}. A displacement occurs when an atom is knocked from its lattice site, a transmutation occurs when an atom absorbs the displacing neutron as well as decaying alpha particles. This damage is augmented in the presence of thermal neutrons. The current investigated spacer was irradiated in a reactor over 14 effective full power years to a damage dose of 30 displacements per atom (dpa). Cross-section of the spring wire was 0.7 mm x 0.7 mm. Samples were cut from different locations, 12 o’clock ($>$300 $^\circ{}C$) and 6 o’clock (180 $^\circ{}C$) respectively along the cross-section. The positioning alters the microstructure as the 12 o'clock segments are not in compression and are at a constant high temperature, while the 6 o'clock position is compressed under the channel with a larger temperature gradient. To ensure appropriate imaging of bubbles, thin samples must be prepared to minimize the number of overlapping bubbles. TEM samples were milled using a focused ion beam (FIB) and then ion polished using a nano-mill with 900 eV Argon ions at $\pm$10 degree glancing angles. All TEM imaging was performed using a JEOL F200 TEM at an operating voltage of 200 keV utilizing single and double tilt sample holders. Two-beam dynamical bright field and weak beam dark field conditions were applied for imaging irradiation induced microstructural changes.

\begin{table}[]
\caption{Chemical composition of X-750 spacers used in CANDU fuel channels.}

\label{tab:comp}
\begin{tabular}{|c|c|}
\hline
Element  & wt. \% \\ \hline
Nickel   & 68.6   \\ \hline
Chromium & 16.0   \\ \hline
Iron     & 8.0    \\ \hline
Titanium & 2.5    \\ \hline
Niobium  & 1.0    \\ \hline
Cobalt   & 1.0    \\ \hline
Trace    & 2.9    \\ \hline
\end{tabular}
\end{table}

\subsection{Data Set Preparation}

Helium generated during neutron irradiation coalesced and formed bubbles, which form roughly spherical defects \cite{zhang2013microstructural}. In the dataset generated by CNL, all features of interest were considered circular, a result of the hydrostatic pressure exerted by the helium gas. The coordinates of a bounding box identifying these circles were used to store the positions and size of the bubbles. Note that this strategy could be extended to handle ellipsoidal defects. The manual analysis involved clicking on the centroid of the bubble and then subsequently clicking on the edge of the bubble. This process is performed twice per bubble to generate an average center and fringe. The bounding boxes used to train the R-CNN are based on these average values.

The open source program LabelImg was used to annotate the images \cite{labelIMG}. The bounding boxes were translated into the XML files to be read by TensorFlow. Micrographs contained upwards of 50 helium bubbles, which were manually quantified by an expert in the bubble classification procedures. Differentiating the smaller bubbles from the base material is challenging, leading to variance in classifications based on interpretation \cite{van2018intra}. The training and validation data was classified by a single trained individual, which minimizes this variance. The training dataset consisted of 230, 512$\times$512 pixel, gray-scale images and being comprised of both over and under-focused images (80 over-focused micrographs and 150 under-focused micrographs). Over and under focused imaging conditions are used to generate the required contrast. The difference in quantity of over and under-focused micrographs is due to a human factor. The scientist analyzing the data would scroll through focal series, starting with under-focused micrographs, and use what they judged to be the highest quality image to proceed with their analysis. The under-focused images tended to be selected more often for full analysis. The over-focused images would occasionally be used as a means to validate the count in the under-focused condition. The micrographs used for training had a fixed 0.38 nm/pixel magnification. 

A test dataset was compiled using a separate collection of micrographs. This test dataset was collected independently from the training dataset, it consists of 23 images. 4 of these 23 images were annotated by three separate experts, and the additional 19 were quantified by one of the experts. These 23 images were taken in over-focus, under-focus, and low magnification conditions. The high-resolution test images had the same magnification as the training set (0.38 nm/pixel) while the low magnification images are captured at $>$1 nm/pixel. These images were used after training to establish model metrics which could be used to validate performance.

\subsection{Model Training}

 As explained in the section \ref{sec:NetArch}, Faster R-CNN behaves as a unified network when training. The training scheme alternates between fine tuning the region proposal network and fine tuning for object detection while keeping the region proposals fixed \cite{NIPS2015_5638}. The model was trained for a period of 4 hours with a standard loss stop being used to prevent over-fitting. In this time 20,000 training epochs were performed. The image dataset consisted of the 230 images (80 over/150 under-focused) of consistent size (512x512) and magnification (0.38 nm/pixel) with each image containing 50-100 identified bubbles. To segment the images and generate the training and validation image sets used for model training, the total dataset was randomly segmented 70/30 with 70\% of the dataset images used to train the model weightings and the other 30\% used to validate those weightings. Following the literature, a stochastic gradient decent method with back-propagation was used to perform the end-to end model training  \cite{NIPS2015_5638}. During training, a mini-batch size of 6 was used which allowed for quick convergence of the loss function and is consistent with literature values for appropriate batch sizes \cite{masters2018revisiting}. The network was trained using Compute Canada servers with a single Intel E5-2683 v4 "Broadwell" Processor clocked at 2.1 Ghz and a single NVIDIA P100 Pascal GPU. A summary of all hyper-parameters tuned for model adaptation to this defect type are available in Table \ref{tab:param}.
 
The prediction of a bounding box yields an objectness score. This value is a measure of the model's confidence that an object is detected. The value which dictates if a detected objects score is deemed true, is given by the confidence threshold. A higher confidence threshold will lead to an increase in precision, at the expense of a decrease in recall (these metrics are defined in the following subsection). The model was set to have a confidence threshold of 0.50. The value of 0.50 leads to high precision ($>$0.90) in target image groups while maintaining reasonable recall ($>$0.70).

\subsection{Model Performance Metrics}
Within this work we report three performance metrics: recall, precision, and $F_1$ Score. These metrics are based upon the intersection over union (IoU) of the predicted bounding boxes with the ground truth values as defined by the manual detection method. IoU is a metric that quantifies the amount of overlap between ground truth bounding boxes and the predicted box \cite{Rezatofighi_2019_CVPR}. This calculated overlap value is used to classify detection boxes into three classes, a true positive (TP), false positive (FP), and false negative (FN). True positives are instances of the model detecting an object that is present, a false positive is when the model detects an object that is not present, and a false negative is when an object is present but is not detected. These values; IoU, TP, FP, and FN, are used to develop the two model performance metrics that are being utilized.

Recall is the true positive rate and it measures the probability of ground truth objects being correctly detected. Recall ranges from 0 to 1 and a value of 0.6 implies that the model correctly predicts 60\% of the objects. The formula for recall is:
\begin{equation}
Recall = \frac{TP}{TP+FN}
\end{equation}
Precision is the probability that the predicted bounding boxes match the actual ground truth boxes. The precision scores also range from 0 to 1 and a value of 0.8 implies that when the model detects an object 80\% of the time it is correct. The formula for precision is:
\begin{equation}
Precision = \frac{TP}{TP+FP}
\end{equation}
In a statistical analysis the $F_1$ Score is a measure of a test's accuracy. The score considers both recall and precision to compute its value, $F_1$ score is the harmonic mean of the recall and precision metrics and varies between 0 and 1. The formula for the $F_1$ Score is:
\begin{equation}
F_1 = 2\cdot \frac{precision\cdot recall}{precision+recall}
\end{equation}

\section{Results}

\subsection{Visual Defect Assessment}
\begin{figure}[hbt!]
\centering
\includegraphics[width=7cm]{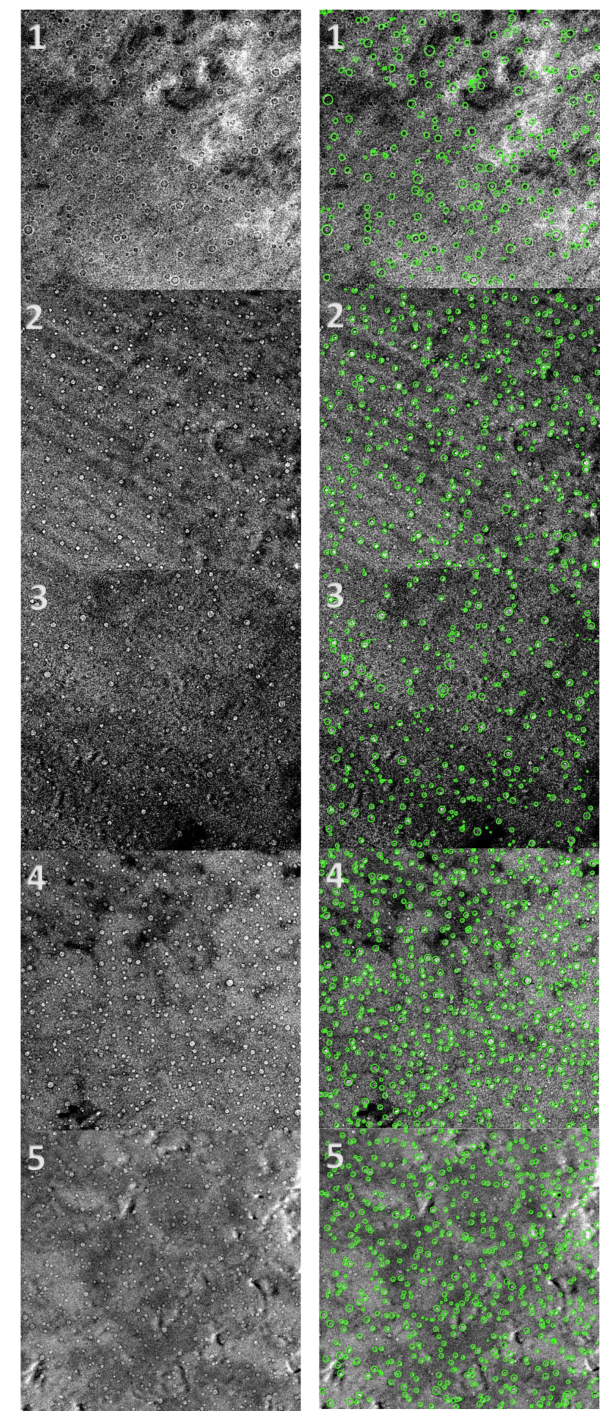}
\caption{Original TEM micrographs are in the left-side panels. The micrographs annotated by R-CNN are in the right-side panels. 1) is an example of an over-focused image. Note that the majority of the training set is comprised of under-focused micrographs. 2,4,5) are representative examples of relatively clean and visible micrographs. 3) is an example of a noisy micrograph.}
\label{fig:annotated}
\end{figure}

Visual examination of bubbles is performed using Fresnel contrast imaging to make the bubbles identifiable in the micrographs. The imaging can be performed either under- or over-focused, which will produce white bubbles against a dark background or black against a light background respectively. The contrast is generated to differentiate bubbles from background noise. Typical examples of over-focused bubbles are shown in Fig. \ref{fig:annotated} image 1, while images 2-5 are under-focused. A dominant defect type is observed: helium bubbles, i.e. spherical cavities where helium produced by neutron interactions accumulate. As illustrated in Fig. \ref{fig:annotated}, the bubbles are of low contrast relative to the background. Additionally, bubble sizes vary between 1-12 nm. As the bubbles decrease in size the contrast diminishes, which further impedes consistent bubble identification. This is especially notable in the presence of FIB damage.

\begin{figure}[hbt!]
\centering
\includegraphics[width=8.6cm]{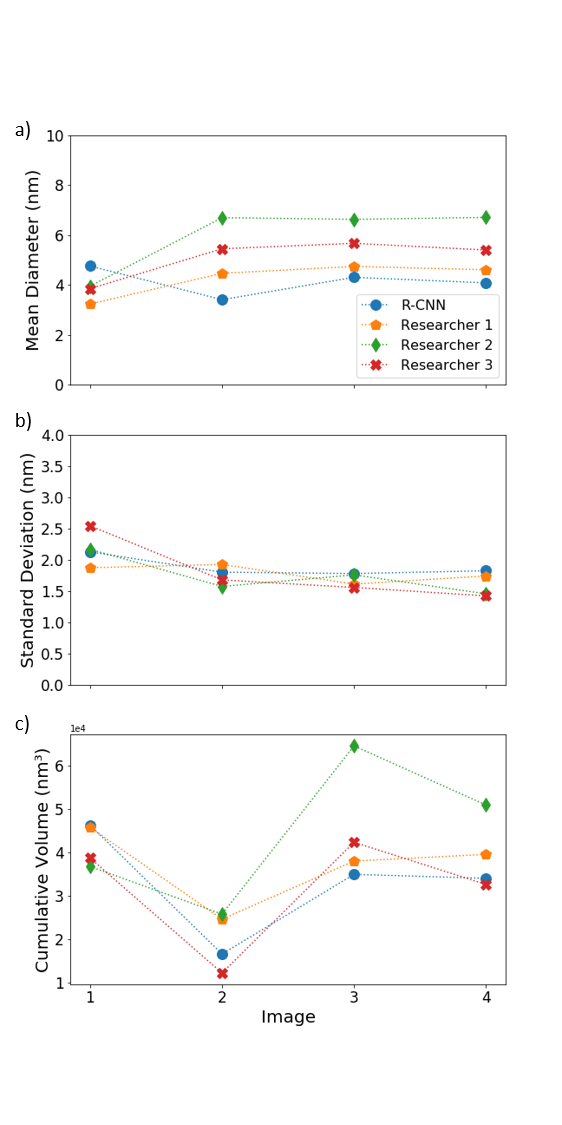}
\caption{a) The mean of the helium bubbles radius distribution b) The standard deviation of the bubble radius distribution c) Total helium bubble volume in four micrographs of irradiated X-750. Each micrograph is quantified by three researchers and the R-CNN}
\label{fig:method}
\end{figure}

\begin{figure}[hbt!]
\centering
\includegraphics[width=8.6cm]{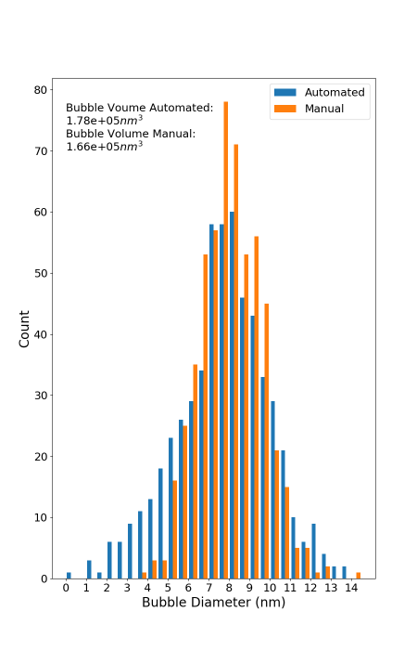}

\caption{A typical example of the bubble size distribution in a low-resolution image, 1 nm/pixel. Both models yield similar uni-modal distribution methods. Both methods yield near identical cumulative bubble volumes.}
\label{fig:hist}
\end{figure}

\begin{figure}[hbt!]
\centering
\includegraphics[width=8cm]{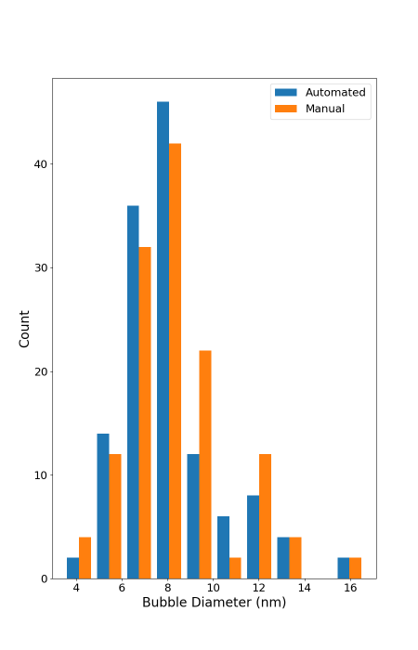}

\caption{A typical example of the bubble size distribution in a high-resolution image, 0.38 nm/pixel. Both models yield similar uni-modal distribution methods.}
\label{fig:hist1}
\end{figure}

As quantification of helium bubbles is an ambiguous task, the statistics generated by visual inspection vary from one human inspector to another. Figure \ref{fig:method} illustrates these variations across four images analyzed by three scientists and the trained R-CNN. The variation is consistent with human analysis performed in other studies \cite{li2018automated, roberts2019deep}. Variations of approximately 25\% are observed from individual-to-individual. The R-CNN-extracted values are not statistically different from those found by the three human scientists. Bubble size distributions can also be used to compare human-based and R-CNN detection. Fig. \ref{fig:hist} represents the distribution as quantified by the detection model and a human at low magnification. Both models exhibit a uni-modal distribution with a peak between 7.5 and 8 nanometer diameters. These distributions generate cumulative bubble volumes that are within 99\% agreement. The primary differences between the methods are a higher peak value with narrower tails on the manual quantification method relative to the automated method. A high magnification distribution is shown in Fig. \ref{fig:hist1} highlighting the high degree of correlation.

\subsection{Performance Metrics}

In gauging the effect that increased training data has on model performance metrics, an iterative training procedure was used. When training with incremental quantities of the cumulative training dataset, improvements in the model performance metrics were observed. Table \ref{tab:train} provides a breakdown of the metric improvements with increasing dataset size when using an IoU of 0.5. There do not appear to be diminishing returns when using a dataset of this size. 

\begin{table}[hbt!]

\caption{Model performance metrics on the test dataset when training with increased quantities of training data. IoU of 0.50 used when generating metrics. Model saturation has not been reached with datasets of this size.}
\label{tab:train}
\centering
\begin{tabular}{|c|ccc|}
\hline
\multicolumn{1}{|c|}{Iteration} & Recall & Precision & $F_1$ Score\\ \hline
60 Images                           & 0.62   & 0.76  & 0.69   \\ \cline{1-1}
120 Images                           & 0.67   & 0.78 & 0.73    \\ \cline{1-1}
180 Images                           & 0.70   & 0.81 &  0.76   \\ \cline{1-1}
230 Images                           & 0.72   & 0.84  & 0.78   \\ \hline

\end{tabular}
\end{table}

Bounding boxes detected by the model in the 23 validation images were compared against human-generated values using IoU to identify the TP, FP, and FN bounding boxes. The comparison was performed to generate the precision, recall, and $F_1$ Scores at varying IoU thresholds. IoU was varied from 0.5 to 0.9, the variation was performed as 0.5 is generally accepted as the minimum overlap needed for a bounding box to be acceptable as a TP with quality of predictions increasing with increasing IoU \cite{Rezatofighi_2019_CVPR}. Table \ref{tab:P&R} provides a breakdown of the performance metrics averaged across the 23 images at increasing IoU thresholds. As IoU increased the corresponding performance metrics decrease. A fairly consistent decline in performance is seen with each increment of 0.1 until a threshold of 0.9 is reached where a substantial decrease is observed.

\begin{table}[hbt!]

\caption{Accuracy of the R-CNN's He bubble analysis as a function of IoU value. As the IoU is increased from 0.5 to 0.9 the corresponding model performance metrics are seen to degrade.}
\label{tab:P&R}
\centering
\begin{tabular}{|c|c|c|c|c|c|}
\hline
          & \multicolumn{5}{c|}{IoU}         \\ \hline
Metric    & 0.5  & 0.6  & 0.7  & 0.8  & 0.9  \\ \hline
Recall    & 0.72 & 0.66 & 0.63 & 0.59 & 0.42 \\ \cline{1-1}
Precision & 0.84 & 0.77 & 0.74 & 0.69 & 0.47 \\ \cline{1-1}
$F_1$ Score        & 0.78 & 0.71 & 0.68 & 0.64 & 0.44 \\ \hline
\end{tabular}
\end{table}

Recall and precision metrics were calculated using the 23 validation micrographs at varying IoU thresholds. The validation images contain samples imaged under different conditions and as such the average performance does not accurately represent each imaging condition. To understand the effect imaging conditions have on model performance, the images are segmented by condition and metrics for each are recorded. This breakdown was performed using an IoU threshold of 0.6. The results are summarized in Table \ref{tab:validation}.  The breakdown strongly indicates that the imaging condition is highly influential in the accuracy of the model with particular emphasis placed on the magnification levels of the images. Over and under-focused images exhibit strong recall and precision figures with under-focus being superior. Low magnification images, where bubble sizes are upwards of three times smaller than other image examples, exhibit poor performance at this threshold.

\begin{table}[hbt!]
\caption{Recall, precision, and $F_1$ Score metrics when segmenting the test dataset by image type using an IoU of 0.6. Training was performed with the full set of 230 images. Performance in the over-focused and under-focused sets taken at high magnification is superior to the poor performance of images taken at lower magnification levels.}
\label{tab:validation}
\centering
\begin{tabular}{|c|ccc|}
\hline
\multicolumn{1}{|c|}{Image Type} & Recall & Precision & $F_1$ Score \\ \hline
Over-focused                           & 0.68   & 0.93  & 0.79   \\ \cline{1-1}
Under-focused                           & 0.89   & 0.96 & 0.93     \\ \cline{1-1}
Low Magnification                           & 0.10   & 0.11 & 0.11     \\ \cline{1-1}
Complete Test Dataset (23 Images)                           & 0.66   & 0.77 & 0.71     \\ \hline
\end{tabular}
\end{table}

\subsection{Comparison with Human Analysis}

\begin{figure}[hbt!]
\centering
\includegraphics[width=8.6cm]{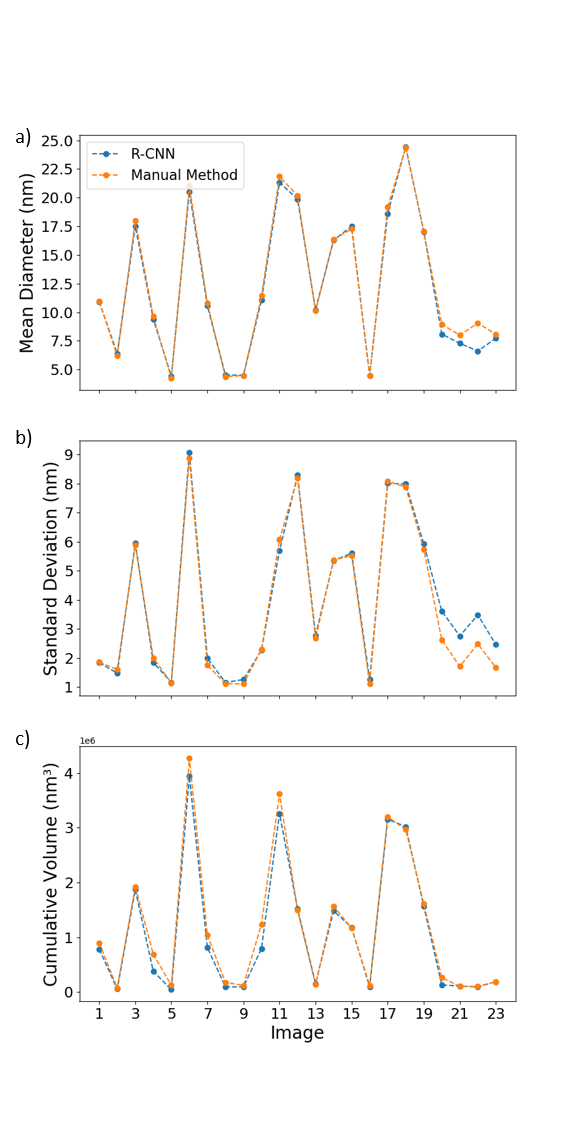}
\caption{a) Mean bubble diameter b) standard deviation of bubble diameter c) cumulative bubble volumes as recorded by the R-CNN and manual procedures across 23 independent images. Images 1-13 are examples of performance in under-focused images, images 14-18 are examples of over-focus, and images 19-23 are of lower magnification.}
\label{fig:ann}
\end{figure}

The bubble statistics for the 23 validation micrographs are plotted in Fig. \ref{fig:ann}. R-CNN and manual results are reported, the validation set contains over-focused images, under-focused images, and lower magnification images. Fig. \ref{fig:ann} shows that the R-CNN and human-generated statistics follow the same trends. The R-CNN and human-based estimates of mean bubble diameter in the higher magnification images labelled 1-18 are within 1.3\% of each other. In the five low magnification images, labelled 19-23, the values are within 12.5\% of each other. The estimates of standard deviations of bubble diameters are within 1.8\% of each other in the high-magnification images, and within 46\% of each other in the low-magnification images. The estimates of total volumes are within 15\% of each other, in both high- and low-magnification images. Bubble area densities can be derived from the cumulative bubble volume and pixel/nm conversion factor of the images, images 1-18 are identical in image area and the bubble volume trends as detected in manual quantification follow that of the automated detection. The R-CNN took approximately 2 seconds to process each image, this is contrasted with the manual quantification procedures which took up to 5 hours per image, this number varies widely on a case-by-case basis. This time savings represent an improvement of four orders of magnitude.

\section{Discussion}

Five indicators suggest that Faster R-CNN is well suited to identifying helium bubbles. First, visual inspections of the five images presented in Fig. \ref{fig:annotated} suggests that the R-CNN accurately identified most of the important features. Second, a comparison between quantitative analyses generated by the R-CNN and a group of human researchers presented in Fig. \ref{fig:method}, involving four images, shows no statistically significant differences.  Third, Fig. \ref{fig:hist} and Fig. \ref{fig:hist1} show that both the human-based analysis and the R-CNN's indicate that the helium bubble follow a uni-modal size distribution in low and high magnification images. Fourth, Table \ref{tab:validation} shows good recall and precision statistics in under-focused and over-focused images . Fifth, Fig. \ref{fig:ann} shows that the R-CNN and human analyses of 23 validation images lead to quantitatively consistent bubble statistics. 

While these results are very encouraging, there are still limitations to the method presented in this article. Performance of the R-CNN was poor when considering images with lower magnifications relative to those in the training set. Likewise, the performance when analysing over-focused images is not as good as that when analysing under-focused images. Table \ref{tab:validation} suggests that adding over-focused images to the training set could improve the model's performance. Adding lower resolution images may help as well. Table \ref{tab:train} suggests that increasing the overall quantity of training data will lead to a corresponding improvement in model performance. 

Substantial correlation between the R-CNN and manual annotation is clearly illustrated in Fig. \ref{fig:ann}, across varying imaging conditions, with the exception of the low magnification condition. In these micrographs, the R-CNN overestimates the number of small-sized bubbles, this is exemplified in Fig. \ref{fig:hist}. There is a smaller than average size and larger standard deviation seen in the low magnification image condition, Fig. \ref{fig:hist} confirms this and shows that the R-CNN is capturing many small objects that are not bubbles. This effect is not seen in high magnification images as shown in Fig. \ref{fig:hist1}. The lower magnification of the sample, and consequently its resolution, is the likely source of error in the underestimation of bubble size. 

It is typically considered best practice to report a measurement error of bubble size, based on the fringe thickness. Our work suggests that such estimates provide a false sense of accuracy: the person-to-person variation in bubble count, Fig. \ref{fig:method}, arguably a more robust estimate of measurement error, is much larger than the typical ratio of fringe thickness to bubble radius. This raises another question: How can automated image analysis provide such error estimates? A possible solution is to independently train different neural networks, based on different architectures and training sets annotated by different scientists. These different networks can then by applied to the same data, and used to estimate measurement errors. 

Analysis of helium bubbles in X-750 is hindered by the inconsistent manner in which the samples are prepared and imaged.  Variation and inconsistencies in training data lead to difficulties in developing the necessary correlations to effectively identify defects. Here we discuss two preponderant issues. First, during sample preparation, if the samples are not thin enough, overlapping bubbles tend to dominate the image; features cannot be consistently identified. Second, when there is excess FIB damage, imperfections appear on the surface. These imperfections appear similar to bubbles and can lead to overestimation of total bubble density. These issues affect both manual and automatic image processing. In addition, junior researchers might not have the experience to recognize that they are dealing with a low-quality sample and therefore not producing useful images. A means of mitigating against spending time on images that cannot produce useful information, in addition to improving training and knowledge transfer between TEM technicians and the scientists, would be to develop a NN able to identify high- and low-quality samples, and give suggestions about possible causes for poor image quality. This could serve as a training tool to help generate large quantities of high and low quality micrographs. 

Recently, it was shown that purpose-built networks are well-suited to identify dislocation-type defects \cite{li2018automated, roberts2019deep}, notably in the context of radiation-induced damage. Our work suggests that an "off-the-shelve" NN, implemented in a popular open-source platform, TensorFlow, is also well-suited for the analysis of micrographs of radiation-induced damage. Our work highlights the viability of adapting an existing network. 

As the field progresses and new features are added, we should mention two challenges. The first is the reliance of R-CNNs on large data sets. The large sets are not always available for materials science applications \cite{agrawal2019deep}. In our case, it is likely that adding images to the training set, notably over-focused and low-resolution images, would improve the model's performance. Note that in order to generate additional training images rapidly, one could use images pre-processed by a preliminary R-CNN, and then manually remove false-positives and add in the false negatives. A second challenge pertains to the automated extraction of defect contours \cite{li2018automated}. This is not an issue when dealing with bubbles since they are largely spherical, but would be an issue for other defect types where such simple shapes are not reliably present. If this model were to be adapted for different defect types it would be recommended to utilize a more advanced segmentation method to extract quantifiable features. 

\section{Conclusions}

The Faster R-CNN was adapted and trained in order to analyze large sets of noisy TEM-generated micrographs. The Faster R-CNN can identify helium bubbles in X-750 alloys after neutron irradiation. It results in bubble statistics in quantitative agreement with those extracted by human-based analysis of the micrographs. The Faster R-CNN can process images in a few seconds, which is orders of magnitude faster than human-based analysis. Accuracy levels of 93\% have been achieved when mapping micrographs imaged at high magnifications and reasonable accurate quantification of samples images at lower magnification. Consistency of imaging conditions is key to the success of the model. Images to be analyzed should be similar to the images used in model training; in particular, focusing conditions should be the same. Manual post processing of Faster R-CNN annotated micrographs can be used to progressively improve the training set, with a lower time investment than full manual annotation. Deep learning shows promise, and can likely be used to improve many other aspects of characterization of materials, including those used for nuclear power generation.

\section{Acknowledgements}

We would like to thank the researchers at the Canadian Nuclear Labs who provided the micrographs used for training as well as the background information on classification of the bubbles. We thank Compute Canada for generous allocation of computer resources. Research was funded by the Canadian Nuclear Laboratories, Mitacs Canada, and the Natural Sciences and Engineering Research Council of Canada.

\section{Data Availability}

The dataset generated for use in this work and analysis, presented in the current study are available for download as supplementary material.

\section{TensorFlow Hyper-Parameters}

Hyper-parameters used in the model training process are shown in Table \ref{tab:param}. The use of these hyper-parameters allows for a replication of the training procedures used in this work.

\begin{table}[h]
\centering
\caption{Hyper-parameters used in the model training configuration file for training the Faster R-CNN}
\label{tab:param}
\begin{tabular}{|c|c|}
\hline
Parameter       & Value  \\ \hline
Learning Rate   & 0.0003 \\ \hline
Mini-batch Size & 6      \\ \hline
Epochs          & 20,000 \\ \hline
Momentum        & 0.9    \\ \hline
Kernel Size     & 16     \\ \hline
\end{tabular}%
\end{table}

\bibliographystyle{unsrt}
\bibliography{scholar}
\pagebreak

\end{document}